\documentclass[aps, prd, twocolumn, lengthcheck, superscriptaddress, 
nofootinbib]{revtex4-1}

\usepackage{epsfig}
\usepackage[usenames]{color}
\usepackage{graphicx}
\usepackage{amsmath}
\usepackage{epstopdf}

\def\bi{\bibitem}

\def\la{\langle}\def\ra{\rangle}
\def\be{\begin{eqnarray}}\def\ee{\end{eqnarray}}
\def\lsim{\mathrel{\rlap{\lower3pt\hbox{\hskip1pt$\sim$}}
     \raise1pt\hbox{$<$}}} 
\def\gsim{\mathrel{\rlap{\lower3pt\hbox{\hskip1pt$\sim$}}
     \raise1pt\hbox{$>$}}} 

\allowdisplaybreaks

\begin{document}

\title{Multifarious roles of hidden chiral-scale symmetry:\\  ``Quenching" ${g_A}$ in nuclei}

\author{Mannque Rho}
\email{mannque.rho@ipht.fr}
\affiliation{Institut de Physique Th\'eorique, Universit\'e Paris-Saclay, CNRS, CEA,  91191, Gif-sur-Yvette, France }

\date{\today}

\begin{abstract}
I discuss how the axial current coupling constant $g_A$  renormalized in scale symmetric chiral EFT defined at a chiral matching scale impacts on the axial current matrix elements on beta decays in nuclei with and without neutrinos. The ``quenched" $g_A$ observed in nuclear superallowed Gamow-Teller transitions, a long-standing puzzle in nuclear physics, is shown  to encode the emergence of chiral-scale symmetry hidden in QCD in the vacuum. This enables one to explore how trace-anomaly-induced scale symmetry breaking enters in the renormalized $g_A$ in nuclei applicable to certain non-unique forbidden processes involved in neutrinoless double beta decays.  A parallel is made between the roles of chiral-scale symmetry in quenching $g_A$ in highly dense medium and in hadron-quark continuity in the EoS of dense matter in massive compact stars.   A systematic chiral-scale EFT, presently lacking in nuclear theory and potentially crucial for the future progress, is suggested as a challenge in the field.
\end{abstract}

\maketitle

\section{Introduction}
It has been recently argued~\cite{quenchedgA,WS}  that the long-standing mystery of ``quenched"\footnote{I put the quotation mark here because what's referred to in the nuclear physics literature as ``quenched $g_A$" is a misnomer. The coupling constant $g_A$ appearing in the nucleonic axial current used to calculate nuclear Gamow-Teller matrix elements is in fact {\it not}  quenched in the sense used in shell-model calculations in the literature.   This will be explained in what follows.  I will however continue the discussions without the quotation mark unless otherwise noted.}  $g_A$ in nuclear super-allowed Gamow-Teller transitions can be resolved by the combination of two mechanisms: one,  strong nuclear correlations controlled by an emerging scale symmetry hidden in QCD and the other, an effect of quantum anomaly in scale symmetry inherited from QCD at the chiral scale $\Lambda_\chi \sim 4\pi f_\pi$ that defines effective field theory for nuclear dynamics.

In this note, I discuss how what triggers the quenched superallowed Gamow-Teller transitions in nuclear medium encodes the emergence of chiral-scale symmetry, hidden in QCD, in strong nuclear correlations and suggest in what way the anomaly-induced breaking of scale symmetry -- referred to by the acronym AISB -- affects how the axial-current coupling constant $g_A$ can indeed be ``fundamentally" renormalized by the vacuum change -- as opposed to the effect of nuclear correlations. This AISB is argued to have an important impact on neutrinoless double beta decay matrix elements.

\section{Superallowed Gamow-Teller Transitions}
To zero-in on essentials for the quenched $g_A$ problem, first consider the superallowed Gamow-Teller decay of the doubly magic nucleus  $^{100}$Sn with 50 neutrons and 50 protons. I pick this case because there is what is claimed to be an ``accurate" data and equally importantly it offers a well-defined theoretical framework. This process allows to exploit the ``extreme single-particle shell model (ESPSM)" description~\cite{ESPSM,RIKEN}. 
In the ESPSM description, the GT process involves, via the spin-isospin flip, the decay of a proton  in a completely filled shell $g_{9/2}$ to a neutron in an empty shell $g_{7/2}$, which can be equated nearly precisely to the GT transition of a quasi-proton to a quasi-neutron on top of the Fermi sea formulated in the Landau Fermi-liquid fixed point theory~\cite{quenchedgA}. Such a feat is not usually feasible for an ESPSM description of generic (say, non-doubly-magic) nuclei. 

Let us begin by defining  the {\it quenching factor}  $q$ often used in the literature
\be
M_{\rm GT} &=& g_A  q {\mathcal M}_{\sigma\tau}.\label{MGT} 
\ee
Here $g_A$ is the free-space (neutron decay) axial-vector coupling constant $g_A= 1.276(4)$ and ${\mathcal M}_{\sigma\tau}$ is the proton-to-neutron single-particle GT matrix element.  The quantity on the right-hand side representing nature should of course be model-independent but $q$ and ${\mathcal M}$ will depend on how ${\mathcal M}$ is to be calculated, so separately model-dependent. The approach adopted in \cite{quenchedgA} gives a precise meaning to what $q$ is in the scheme and how it can be related to the experimental value. As emphasized there, given the superallowed transition with zero momentum transfer,  {\it the quantity ${\mathcal M}$ is just the spin-isospin factor for the Fermi-liquid as well as ESPSM descriptions with all the interaction effects, fundamental (i.e., AISB) as well as of pure nuclear correlations, lumped into the ``quenching" factor $q$.}  In general the two are of course intricately mixed and it is difficult, if not impossible,  to disentangle them in nuclear processes. However if one assumes that the AISB effect is small, then the axial-current can be written with the anomaly factor $q_{\rm ssb}$ representing the scale symmetry breaking (adopting the acronym ``ssb") simply multiplying the scale-invariant axial current as
\be
q_{\rm ssb} g_A \bar{\psi}\tau^\pm\gamma_\mu\gamma_5\psi\label{qssb}
\ee
with 
\be
q_{\rm ssb}= c_A+(1-c_A)\Phi^{\beta^\prime}.\label{ssb}
\ee  
Here $c_A$ is a (in general) density-dependent constant and $\beta^\prime$ is the anomalous dimension of the gluon stress tensor ${\rm tr} G_{\mu\nu}^2$, both of which remain, up-to-date, unknown for QCD with the flavor number $N_f\lsim 3$.   
The quantity $\Phi$ is the ratio $f_\pi^\ast/f_\pi$ with the $\ast$ indicating density dependence of the nuclear medium.  $\Phi$ has been measured by deeply bound pionic nuclei~\cite{kienle-yamazaki}, so is known up to the nuclear matter density $n_0$. Note that because of $\Phi$, $q_{ssb}$ is explicitly density-dependent. In the vacuum, $\Phi=1$, so there will be no dependence on $\beta^\prime$.  Furthermore if $c_A$ were equal to 1 either on symmetry grounds or by density effects in medium, there would be no dependence on $\beta^\prime$. In both cases, the impact of scale symmetry breaking will be absent in the GT transitions.

In \cite{quenchedgA}, resorting to an EFT that incorporates both hidden local symmetry and hidden scale symmetry in chiral Lagrangian,  the matrix element of the {\it superallowed} GT matrix element  was calculated in the large $N_c$ and large $\bar{N}$ approximations in the Fermi-liquid approach. In QCD, the $g_A$ is proportional to $N_c$ in the large $N_c$ counting and in the Fermi-liquid  theory, the Fermi-liquid fixed point is given by $O(1)$ term in the limit $\bar{N}=k_F/(\Lambda_{\rm FL}-k_F)\to \infty$ with $\Lambda_{\rm FL}$  the cut-off in the Fermi-liquid renormalization group decimation.  In these double limits, it comes out that
\be
\la \bar{\psi}\tau^\pm\gamma_\mu\gamma_5\psi\ra_{fi}=q^{\rm Landau}_{\rm snc} \la {\tau^\pm\sigma}\ra_{fi}
\ee
where the entire strong nuclear correlation (``snc") effects are captured in
\be
q_{\rm snc}^{\rm Landau}=(1-\kappa)^{-2}\label{snc}
\ee
with
\be
\kappa= \frac 13 \Phi \tilde{F}_1^\pi.
\ee
Here $\tilde{F}_1^\pi$ is the pionic contribution to the Landau mass\footnote{The pionic contribution to the Landau mass is the Fock term, which is formally of $O(1/\bar{N})$, that goes beyond the Landau Fermi-liquid fixed point approximation made in (\ref{snc}). However soft theorems for both chiral symmetry (pion) and scale symmetry (dialton) figure for the validity of the relation (\ref{snc}), so the pionic contribution is essential. See \cite{WS} for discussions on this matter.},  governed by chiral symmetry for any density $\leq n_0\approx 0.16$ fm$^{-3}$. Therefore the quantity $\kappa$  is almost completely controlled by low-energy theorems. In addition, due to near compensation of the density effects in the two factors, $\kappa$ remains remarkably insensitive to density. It varies negligibly between $n_0$ and $n_0/2$. Therefor it is applicable equally well to both light and heavy nuclei. 

From (\ref{ssb}) and (\ref{snc}), we get the total quenching factor
\be
q=q_{\rm ssb} q_{\rm snc}^{\rm Landau}.\label{qfactor}
\ee

To see what we have, let's first ignore the AISB and set $q_{\rm ssb}=1$. Evaluating $q_{\rm snc}$ in the Landau Fermi-liquid theory, 
we get at $n\approx n_0$~\cite{quenchedgA} 
\be
q_{\rm snc}^{\rm Landau}\approx 0.79.\label{q_L}
\ee
 Thus what is identified as the ``scale-symmetric effective $g_A$"  given in the Landau Fermi-liquid theory is
\be
g_A^{\rm ss} \equiv g_A^{\rm Landau}\approx g_A q_{\rm snc}^{\rm Landau} =1.276 \times 0.79\approx 1.0.\label{gA1}
\ee
Note that $g_A^{\rm ss}$ is {\it not a quenched} coupling constant. The quenching is in the nuclear interactions captured in $q_{snc}^{\rm Landau}$. More on this crucial point below.

\subsection{Fermi-Liquid Fixed Point $\approx$ Extreme Single Particle Shell  Model}  

In what sense can this result (\ref{gA1}) which will figure crucially in what follows  be taken reliable? It relies on taking two limits. One is the large $N_c$ limit in QCD combined with soft theorems. The other is the large $\bar{N}$ limit in the renormalization-group approach to Landau-Fermi liquid theory in treating strongly correlated nuclear interactions. The former is akin to the Goldberger-Treiman relation which is known in the matter-free vacuum to be accurate within a few $\%$ discrepancy. Thus  combining the two limits in nuclear processes can be considered to be comparable to applying to the Goldberger-Treiman relation in nuclear dynamics. It has been shown that the Landau-Fermi liquid fixed point approximation  combined with soft theorems give very accurate predictions for such nuclear electroweak processes as the anomalous orbital gyromagnetic ratio of the proton  $\delta g_l^p$ and the axial charge $0^\pm \leftrightarrow 0^\mp$ $\Delta T=1$ transitions in heavy nuclei~\cite{chiralfilter}.\footnote{To the best of my knowledge, there are no other approaches that match the precision in the predictive power for these quantities.} I would consider (\ref{gA1}) to be of the same accuracy.  At present one cannot make any more precise error assessment.  

As argued above, we can equate $q_{\rm snc}^{\rm Landau}$ to the ESPSM description of the quenching factor of the doubly magic nucleus 
\be
q_{\rm snc}^{\rm Landau}\simeq q_{\rm snc}^{\rm ESPSM}.\label{qLandau}
\ee
Thus the total quenching factor can be written as
\be
q=q_{\rm ssb}q_{\rm snc}^{\rm Landau}\simeq q_{\rm ssb} q_{\rm snc}^{\rm ESPSM}.\label{conjecture}
\ee 

\subsection{Scale-Symmetric Effective $g^{\rm ss}_A$}
That $g_A^{\rm ss}$  is close to 1 raises two questions.  The first is what does the effective $g_A^{\rm ss}$ approaching 1 signify and the second is what does it represent in comparison with experiments?

Since the direct impact of matching to QCD at the given scale $\Lambda_\chi$ is factored out, $q_{snc}$ encodes {\it pure} nuclear correlation effects. Thus $g_A^{\rm ss}$ in (\ref{gA1}) represents their effects captured in the  constant $g_A q_{snc}$ multiplying the non-interacting single particle transition involving only the spin-isospin flip taking place (a) on the Fermi surface for the LHS of (\ref{qLandau})  and (b) in the ESPSM for the RHS.

To put more precisely, $g_A^{\rm ss}$ would correspond to a full-scale shell-model calculation with the {\it unquenched or un-renormalized} $g_A$  that takes into account the excitation of {\it all} configurations  connected to the parent (ground) state by nuclear forces, in particular, the nuclear tensor forces up to excitation energies of $\Delta E\lsim (m_\Delta-m_N)$. Such a full scale microscopic calculation for a mass number of $\sim 100$ would be far from feasible at present, so the relation (\ref{qLandau}) could at best be  conjectural. However  highly powerful numerical techniques have been developed up-to-date that can handle the dynamics of few-nucleon systems, say, $A\lsim 10$. Such a  microscopic numerical calculation in light nuclei anchored on an effective field theory suitably defined at, say, the chiral scale would do the same physics as the mapping of the Fermi-liquid fixed point approximation to the  ESPSM description for many-nucleon systems. Indeed a recent quantum Monte Carlo (QMC) calculation of Gamow-Teller matrix elements entering $\beta$ decay and electron capture rates in $A=3-10$ nuclei was shown, quite convincingly by King et al~\cite{wiringa},  to reproduce the experimental Gamow-Teller matrix elements at a few $\%$  level.  {\it Note that no {\it quenched} coupling constant is involved in this calculation.} There two-body exchange-current operators which enter, chiral-order-suppressed, at N$^3$LO ~\cite{parketal} were included, but found to contribute insignificantly, say, only at the (2-3)$\%$ level as predicted a long time ago~\cite{parketal}. As argued in \cite{quenchedgA}, however, there are no theoretical justifications within the framework of chiral effective field theory to ignore next-order terms of N$^r$LO for $r>3$ if the $r=3$ terms, whether negligible or not, are taken into account. The reason for this is that there can be  non-negligible cancelations between the calculated  $r=3$-order terms and the higher-order terms of comparable strength that are however not calculable due to large number of unknown counter terms.  Therefore  it was argued in \cite{quenchedgA} that the many-body terms be dropped for consistency. Furthermore the possible (2-3)$\%$ contribution can be of the same order as the AISB effect $q_{\rm ssb}$ contributing at density $n< n_0$  as will be made clear below. 

One might think that $g_A^{\rm ss}$ is close to 1 may merely be an accident. Surprisingly however  $g_A^{\rm ss}\approx 1$ is a ubiquitous  and pervasive outcome of simple shell-model calculations in light nuclei (see e.g., \cite{gA-SM}) and has prompted many authors, starting with Wilkinson~\cite{wilkinson}, to ask whether it has something fundamental intrinsic of QCD. Indeed if the Fermi-liquid theory is extended to $\Delta$-hole configurations and the Landau-Migdal Fermi interaction $g_0^\prime$ is assumed to be universal, that is, $g_{NN}^0=g_{N\Delta}^0=g_{0\Delta\Delta}^\prime$, the $\Delta$-hole loop contribution can be made to screen the $g_A$ by the same quenching factor $q\approx 0.80$ as $q_{\rm snc}^{\rm Landau}$.   However this  result would make sense only if all nuclear correlations up to the energy scale of the $\Delta$-N mass difference turned out to sum to nearly zero.  But there is absolutely no reason known why this can be so. Furthermore even if one were to formulate, assuming it makes sense,  the Fermi-liquid theory in the space of both nucleon and $\Delta$ -- Landau-Migdal theory -- there would be no reason why the universality should hold for the $g_0^\prime$ channel.  

In any event  I must say it is highly puzzling  why the {\it naively calculated} $\Delta$-N screening leads to the more or less same quenching as the purely nucleonic Landau Fermi-liquid correlations do -- even if it were coincidental. This has led to a long-standing confusion in the field. This clearly points to that what is ``fundamental" depends on how and at what scale the EFT is matched to QCD.

\subsection{$g_A^{\rm ss}$ and Dilaton-Limit Fixed-Point} 

Let me now address what $g_A^{\rm ss}\approx 1$ could imply. As noted, the axial current $g_A\bar{\psi}\gamma_\mu\gamma_5\psi$ is scale-invariant.  Thus the nuclear interaction that leads to (\ref{qLandau}) reflects scale symmetry because it is associated with both soft-pion and soft-dilaton theorems although the NG boson masses figure in the Landau interactions involved. Again this is quite analogous to the Goldberger-Treiman relation which holds both with and without the pion mass as long as the mass is small compared with the QCD scale, reflecting soft theorems in chiral symmetry. That $g_A^{\rm ss}\approx 1$ is not accidental but could perhaps reflect scale symmetry in action  can be seen in what happens as baryonic matter is tweaked to the putative chiral restoration or more appropriately to the infrared (IR) fixed  point in the ``genuine dilaton" theory of scale symmetry~\cite{CT,crewther}.\footnote{The idea of genuine dilaton may  be controversial in technicolor approaches to the BSM, but we find it highly appropriate for nuclear dynamics as reviewed in \cite{MR-PCM,WS}.} This is because in the chiral limit, approaching the chiral restoration or the IR fixed point, the dilaton condensate  and the quark condensate approach each other,  $f_\chi\to f_\pi$, in  arriving at what is called ``dialton-limit fixed-point (DLFP)"~\cite{bira} relevant for compact-star matter~\cite{MR-PCM}.  In the EFT giving $g_A^{\rm Landau}$, going toward  the DLFP by increasing density is found to impose the constraint on the quark axial current $g_A^{Q}\bar{Q}\gamma_\mu\gamma_5 Q$ -- where $Q$ is the quark field - which is likewise scale-invariant 
\be
g_A ^{Q}\to 1.
\ee
This corresponds to a non-interacting quasi-quark in the medium in which $f_\pi\simeq f_\chi\neq 0$ making the GT transition. It resembles -- uncannily --  the quasi-proton making the superallowed GT transition in the ESPSM in the doubly magic nucleus {\it in the absence of AISB effect}, i.e., with $q_{\rm ssb}=1$. It is thus tempting to consider that {\it albeit} approximate scale symmetry permeates from low density -- nuclei -- to high density -- compact-star matter. This could be considered as one of multifaceted manifestations of hadron-quark continuity discussed in \cite{MR-PCM,WS}. An equally striking manifestation of continuity between hadrons and quarks, which may be related to the pervasive continuity of scale symmetry in $g_A$,   is in the equation of state (EoS)  of dense baryonic matter. Conformal symmetry, broken both by anomaly and spontaneously -- and  invisible in nuclear systems -- emerges pervasively in the  unitarity limit at low density in light nuclei~\cite{unitarity-bira} and in the pseudo-conformal sound speed $v_{sPC}^2/c^2\approx 1/3$ in massive compact stars at high density,  $n\gsim 3n_0$~\cite{WS}, an observation that could be considered as a critical opalescence to the conformal speed $v_c^2/c^2=1/3$ expected at asymptotic density in QCD. 

\section{Strongly Forbidden Axial Transitions}
It follows from the discussions made so far that (\ref{gA1}) is not a {\it fundamental} coupling constant to be applied generically to {\it any} axial responses in the background defined by the modified vacuum. It encodes as completely as feasible nuclear correlations generated in the given background for the superallowed GT transitions in nuclei. It is  not applicable to  nuclear axial responses different from the superallowed (zero-momentum transfer) GT transitions. It is in particular {\it not} applicable to no-nunique strongly forbidden transitions. This should be highly relevant to neutrinoless double beta decay processes~\cite{ejiri} -- important for going  BSM -- where the momentum transfer involved can be of order $\sim 100$ MeV. 

This problem is recently highlighted by Kostensalo et al.~\cite{cobra}  in the ``effective" value of $g_A$,  denoted $\bar{g}_A$,  in the $\beta$ decay  spectrum-shape function involving leptonic phase-space factors and nuclear matrix elements in forbidden non-unique beta decays.  The nuclear operators involved there are non-relativistic momentum-dependent impulse approximation terms. In principle there can be  $n$-body exchange-current corrections with $n \geq 2$. However the corrections to the single-particle (impulse) approximation are typically of $m$-th order with $m \gsim  3$ relative to the leading one-body term in the power counting in $\chi$EFT and could be -- justifiably --   ignored.  Unlike in the superallowed GT transition, neither soft theorems nor the Fermi-liquid renormalization-group strategy can be exploited. Hence (\ref{gA1}) is not relevant to the spectrum-shape function discussed in \cite{cobra}. What is  relevant instead is the scale-anomaly correction (\ref{ssb}).

Given that experimental data are available for the superallowed decay in $^{100}$Sn~\cite{ESPSM,RIKEN}, one can first analyze what one learns from it. The recent RIKEN data~\cite{RIKEN} is claimed  to be a big improvement with much less error bars over what's given, among others, in \cite{ESPSM}, so let me look only at the RIKEN result. As before\footnote{I eschew the kind of ``error analyses" done in standard chiral EFT circles which are not needed for the new ideas developed here. This is because here I am dealing with distinctively qualitative effects, not quantitative. Going beyond the accuracy commensurate with the soft theorems I relied on will require a formulation of high-order chiral-scale perturbation theory much superior in power and quality to what's available up to now.}  I will not quote error bars in arriving at $q_{ssb}$.  From the RIKEN data, dividing by the ESPSM matrix element, the experimental quenching factor comes out to be
\be
q_{\rm riken} ^{\rm ESPSM}\approx 0.50.\label{rikenq}
\ee
Equating  $q_{\rm snc}^{\rm Landau}$ to what is given in ESPSM, i.e., $q_{snc}^{\rm ESPSM}$, and 
dividing  (\ref{rikenq}) by (\ref{q_L}), one obtains
\be
q_{\rm ssb}^{\rm RIKEN} \approx 0.63.\label{rikenssb}
\ee
The deviation $\delta=(1-q_{\rm ssb}^{\rm RIKEN}$) represents the AISB effect.

Let me suppose that the future measurements will confirm the present RIKEN data. Then (\ref{rikenssb}) can be used to deduce the axial coupling constant $\bar{g}_A=g_A q_{\rm ssb}$ in the axial current $\bar{g}_A \bar{\psi}\gamma_\mu\gamma_5\psi$ applicable to {\it all axial transitions in the background of given density}
 \be
\bar{g}_A^{\rm RIKEN}=g_A q_{\rm ssb}^{\rm RIKEN}=1.276\times 0.63\approx 0.80.\label{riken-ssb}
\ee
In distinction to $g_A^{\rm ss}$, this is the ``genuine" {\it quenched} axial vector constant. 
 
Now how does  this prediction compare with the results of the analysis -- referred to as ``COBRA" -- of  the spectrum-shape factor of $^{113}$Cd  $\beta$ decay in \cite{cobra}? Listed in the increasing  average $\chi$-square  are 
\be
\bar{g}_A^{\rm COBRA}
=0.809\pm 0.122,\ 0.893\pm 0.054,\ 0.968\pm 0.056.\nonumber\\
\label{cobraexp}
\ee
Allowing possible uncertainty in (\ref{rikenssb}), it seems reasonable to take these results~\cite{cobra} to be consistent with the prediction (\ref{riken-ssb}).
%

There are however two caveats to keep in mind in assessing the results given above.

The first is the accuracy of  (\ref{rikenssb}) extracted from the RIKEN experiment. In the previous experiments as, e.g., in \cite{ESPSM}, various different values of $q_{\rm ssb}$ were obtained. Thus it is highly desirable  that the RIKEN measurement  be reconfirmed or improved on. It will be seen below that the RIKEN value for $q_{\rm ssb}$ (\ref{rikenssb}) as it stands could get in tension with the theory for the AISB effect as given by Eq. (\ref{ssb}).

The second  is the nuclear model dependence in the analyses leading to (\ref{cobraexp}). Involved is the current operator  $\bar{\psi}\gamma_\mu\gamma_5\psi$ for strongly forbidden $\beta$ decay.  It is known in chiral perturbative approach since a long time~\cite{parketal} -- and reconfirmed in subsequent refined calculations --  that many-body (exchange current) corrections to the single-particle spatial component  of the axial current are suppressed by high chiral orders, at least by 3 orders, so the impulse approximation -- ignoring multi-body currents -- in the axial channel, unless accidentally suppressed, should be sufficiently reliable in dealing with the strongly forbidden transitions involved.\footnote{The story of the unique first-forbidden transition mediated by the axial-charge operator is a totally different story. There two-body exchange charge operator is dominated by terms governed by soft-pion and soft-dilaton theorems as predicted by theory and confirmed by experiments a long time ago~\cite{chiralfilter}. It does not however affect non-unique highly forbidden  operators concerned here.} The operators involved are necessarily non-relativistic and are in principle known up to a manageable form reduced from the covariant current. However the numerical results for $\bar{g}_A$ must sensitively depend on nuclear models that enter in the calculation of the transition matrix elements involved.   It is not easy to assess the quantitative reliability of the nuclear models in giving highly forbidden non-unique transition matrix elements. In particular one cannot achieve for non-unique transitions  the accuracy with which the $q_{snc}$ can be computed in the superallowed kinematics of the doubly magic nucleus. This is because the mapping of the Landau Fermi-liquid fixed-point theory to the ESPSM description exploited in the superallowed GT transition is not feasible for forbidden transition kinematics.

\section{Anomaly-Induced Symmetry Breaking}
Let me now turn to the AISB effect (\ref{ssb}) and how the caveats mentioned above impact on confronting theory with observations. 

The two parameters that figure in the theory are quantities that are inaccessible perturbatively. They are intrinsically nonperturbative and remain un-calculated even in the matter-free vacuum~\cite{CT}. Furthermore they are inaccessible by QCD lattice simulation due to the density dependence in nuclear processes.  The only thing one can say at present is that the ``genuine dilaton" approach~\cite{crewther,CT} requires that $\beta^\prime >0$.  Thus both are as yet unknown parameters. 

Given that there are two unknown  parameters, one cannot make a unique determination of $q_{\rm ssb}$ from only the RIKEN result.  However as discussed in \cite{omega}, approaching dense matter in the skyrmion description in the presence of hidden local symmetry and dilaton scalar mesons -- the key degrees of freedom that figure crucially in the EoS of compact stars in \cite{MR-PCM} --  is found to break down disastrously unless the homogeneous Wess-Zumino (hWZ) Lagrangian $ {\mathcal L}_{\rm hWZ}$  -- which is scale-invariant  like the axial current --  is corrected by an AISB factor of the form
\be
q_{\rm hWZ}=c_{\rm hWZ}+ (1-c_{\rm hWZ}) \Phi^{\beta^\prime}. 
\ee
As with $c_A$, $c_{\rm hWZ}$ is unknown. It turns out -- fortunately -- that if one picks $1 \lsim  \beta^\prime\lsim 3.5$\footnote{The anomalous dimension $\beta^\prime$ has been extensively studied in nonabelian gauge theories for  $N_f\gsim 8$ in connection with the technicolor QCD for going BSM. But there is no information for $N_f\sim 3$ that is concerned here.} and $0\lsim c_{\rm hWZ} \lsim 0.2$,  then one  can not only eliminate the catastrophe but also obtain a qualitatively reasonable phase structure of the dense matter~\cite{omega}.  It turns out that with $c_{\rm hWZ}\approx 0$ and $\beta^\prime\approx (2-3)$, a satisfactory result can be obtained. 

The high density-matter structure addressed in \cite{omega}  is to zero-in on the density regime where the chiral condensate $\la\bar{q}q\ra$ tends toward zero. Here one is dealing with  $\Phi\ll 1$.  The density regime involved in $\beta$ decay is instead near the equilibrium density $n_0$ at which $\Phi$ is $\lsim 1$. In this case it may be unreliable to adopt the value of $\beta^\prime$ appropriate for the regime where $\Phi\ll 1$.

Just to have an idea how the expression (\ref{ssb}) fares, let me simply take $\beta^\prime\approx 2.5$ indicated from the dense matter case of \cite{omega}. Now assuming that $\Phi\approx f_\pi^\ast/f_\pi\approx 0.79$\footnote{This corresponds to density $n\approx n_0$. I am assuming that one can use the nuclear matter value although nuclei of mass number 113 are concerned.}, one can evaluate $c_A$'s corresponding to (\ref{cobraexp})
\be
0.15 \lsim c_A^{exp}  \lsim 0.48.
\ee
  In fact $c_A\approx 0.15$ gives $\bar{g}_A$ compatible with the superallowed RIKEN data as well as with the skyrmion crystal structure for dense matter. 

It turns out however that the density dependence of $\Phi$ in nuclear $\beta$-decay processes can play an extremely important role if  $\beta^\prime$ turns out to be $>1$.  Suppose the nuclei considered in \cite{cobra} have the average density $\sim 0.6n_0$ as in the deeply bound pionic nuclei of mass number $\sim 110$ measured~\cite{kienle-yamazaki}, then only the $q_{\rm ssb}$ that yields $\bar{g}_A\approx 1$ will be acceptable. Others will be ruled out. Furthermore  if $\beta^\prime < 2$,  even the RIKEN data will be in serious tension with the AISB formula (\ref{ssb}). In this case either the RIKEN measurement will be cast in doubt or the AISB formula (\ref{ssb}) relying on linearized approximation will be in error.

\section{Remarks}

While what's discussed in this note  is quite interesting in that nuclear beta decay with strong sensitivity to density can give a glimpse into how hidden scale symmetry can be probed,   a long-standing problem in strong-interaction physics, there are many issues to resolve to make what's obtained here truly sensible. 
First of all there are no known reasons why $c_A$ should be directly related to $c_{\rm hWZ}$  except that they are of the same RG nonperturbative effects and the $c$ coefficients for a given $\beta^\prime >0$ could very well be correlated. It is nonetheless exciting that nuclear processes can offer an {\it albeit}  first hint on  $\beta^\prime$ for $N_f$ as low as 3,   up-to-date a totally unknown quantity in QCD. Secondly it is imperative to check that similar AISB effects do not appreciably upset the properties of normal nuclear matter as well as of  dense matter  that are more or less well described with the $c$ coefficients set equal  to 1, i.e., with no {\it explicit} AISB effects.  So far there seem to be no indications for $c\neq 1$ in the EoS in nuclear physics. If it turned out that  $c_\zeta\approx 1$ for certain channels $\zeta$ and $\approx 0$ for other channels, it would be absolutely necessary to understand how and why. 

What transpires from the result discussed in this note is that scale symmetry combined with chiral symmetry must play a crucial role, as argued in \cite{WS},   in nuclear dynamics ranging from nuclear density all the way to compact-star matter density. The  work available up-to-date in the literature of incorporating scale symmetry with chiral symmetry in nuclear physics is far too involved with too many unknown parameters to be on the right track in view of the power and simplicity found in the large $N_c$ and large $\bar{N}$ limit in resolving the intricate $g_A$ problem. Much work needs to be done.

\subsection*{Acknowledgments}
I am grateful for helpful comments from Jouni Suhonen on work in progress with his collaborators and for on-going discussions/collaborations with Yong-Liang Ma.

\end{document}